# Mean Motion Resonances With Nearby Moons: An Unlikely Origin For The Gaps Observed In The Ring Around The Exoplanet J1407b

Phil J. Sutton*

psutton@lincoln.ac.uk

University of Lincoln, School of Mathematics and Physics, Brayford Pool, Lincoln, LN6 7TS, UK

Corresponding author: Phil Sutton, psutton@lincoln.ac.uk

# Abstract

With the use of numerical models, we investigate whether Mean Motion Resonances (MMR) with nearby moons to the J1407b ring system were the cause of the observed $0.0267 AU$ wide gap located at $0.4 AU$. Only one location of a moon at $0.63 AU$ (corresponding to a 2:1 MMR) was found to form a gap at $0.4 AU$ over short time periods of $< 100 yr$. However, the proximity of a low mass moon ($0.08 M_\oplus$) caused significant scattering of the outer ring edge at $0.6 AU$, along with the formation of an additional gap at the 3:2 MMR ($0.485\ AU$), which is not consistent with observations. Further models with moons located at MMR's 3:1, 4:1, 7:3 and 5:3 failed to form gaps at $0.4 AU$ for time periods $< 100 yr$. Instead, gaps were formed in the ring at 3:2 and 2:1 MMR's which resulted in gaps at radial locations between $0.44 - 0.56 AU$. Additionally, gaps also take longer than one orbital period of J1407b about the primary to form. Given that J1407b is on a highly eccentric orbit and is thought to strongly perturb the ring at apocentre it appears unlikely that gaps form due to MMR's with nearby moons as opposed to embedded moons. Including an appropriate total mass of the ring equal to Earth a dampening effect was witnessed on the gap formation process, causing an increase in the time required to open a gap due to MMR's. Therefore, we conclude the observed gap at $0.4\ AU$ is unlikely to be caused by MMR's with nearby moons.

1. Introduction

The discovery of a potentially massive ring system around the exoplanet J1407b presents new and exciting challenges in the studies of exoring dynamics (Rieder & Kenworthy 2016; Kenworthy & Mamajek 2015; Mamajek et al 2012), with further work underway to discover more Saturnian like ring systems around exoplanets (Aizawa et al 2017; des Etangs et al 2017; Kohler 2017; de Mooij et al 2017; Hatchett et al 2018). J1047b is the first large gas giant (~20 $M_{Jupiter}$) believed to have a large ring outside of our Solar System. A possible solution to the unusual transit of J1407b across the primary was shown to be a large planetary ring that extends $0.6AU$ radially from the planet. This makes the ring system 200 times larger than that of Saturn's which only have a radial extent of $\sim 1 \times 10^{-3} AU$. Planetary ring models with the best fit also hinted towards a radially asymmetric ring with apparent gaps. An embedded moon has been suggested as a potential mechanism for the largest of these inferred gaps at $0.4AU$, which has a width of $0.0267AU$ and is expected to be $\sim M_\oplus$ (Kenworthy & Mamajek 2015). Lindblad resonances with outer moons like those seen in Saturn's rings (Goldreich & Tremaine 1978; Lissauer & Cuzzi 1982) is also a possible solution for gaps within planetary rings. The moon mass ($M_{moon}$) was analytically derived by Kenworthy & Mamajek (2015) assuming an embedded moon would gravitationally clear a gap with width at least equal to its Hill radius ($R_{Hill}$). Therefore, placing an upper limit of $0.8M_\oplus$ on a potential embedded moon.

$$R_{Hill} = a(1-e)\sqrt[3]{M_{moon}/3M_{J1407b}}, \qquad [1]$$

Where $a$ is the semi-major axis and $e$ is the eccentricity.

More advanced detection techniques (Kipping 2009; Pasqua & Assaf 2014; Kipping et al 2015; de Mooij et al 2017; Aizawa et al 2018) are being developed for finding exomoons and exorings, which have the potential to discover more systems like J1407b. Therefore, our understanding of ring-moon interactions derived from the studies of Saturn's rings becomes ever more important. One key point of the hypothesised J10407b exoring is how gravitational interactions of nearby moons can create truncated edges to gaps. In Saturn's A ring the small embedded moons Pan and Daphnis gravitationally clear ring material in the Encke and Keeler gaps. The width of the gap formed for large moons scales with mass such that $D_{gap} \propto M_{moon}^{2/3}$ (Petit & Hénon 1988; Horn et al 1996; Torrey et al 2008; Weiss et al 2009). This only considers a complete gap that spans $2\pi \; radians$ in azimuth. Here, gap widths are larger than the Hill radii of the embedded moons due to gravitational perturbations that exist beyond the Hill radius. Both Pan and Daphnis create gaps at least an order of magnitude greater than their respective Hill radius. Analytically it has been shown Pan and Daphnis create gaps with half widths of $\Delta a \approx 3.8 R_{hill}$ (Weiss et al 2009). In this case it is assumed that $\Delta a$ is small compared to the semi-major axis $a$ of the moon and that ring particles initially encounter the moon on circular orbits. Additionally, gap widths and profiles from embedded moons are derived by balancing the moon's gravitational torque with the viscous torque of the ring (Crida et al 2006; Cuzzi et al 2010). This furthermore suggests that the previously proposed upper mass of $0.8 M_\oplus$ has room for refinement with embedded moons having masses $M_{moon} \ll 0.8 M_\oplus$.

Embedded moons are only one physical mechanism that can create a gap in a planetary ring. Orbital resonances may also exist with larger moons located external to the rings. Here, ring particles experience a gravitational tug at the same location in their orbit when the semi-major axis of ring particles and moon result in integer multiples of their respective orbital periods (Goldreich & Tremaine 1978; Lissauer & Cuzzi 1982), known as mean motion resonances (MMR). The most obvious example in Saturn's rings is the Cassini Division caused by a 2:1 ILR (Inner Lindblad Resonance) with the moon Mimas (Goldreich & Tremaine 1978; Noyelles et al 2016). Saturn's B ring also shows evidence of similar resonances, with the nearby smaller moons Prometheus, Pandora, Atlas and Janus/Epimetheus. Stellar occultations then show the effect these resonances have on the ring which manifest themselves in the form of gaps and density waves (Colwell et al 2007;

Hedman & Nicholson 2016). Spiral density waves in planetary rings from an outer moon, also known as Lindblad resonances, are formed when a particle in the ring has a radial epicyclic frequency integer multiples of the forcing frequency (Nicholson et al 2014; Shu 2016). In this case the perturbing moon.

For considerably smaller moons, or moonlets, previous theoretical work (Lewis & Stewart 2009; Crida et al 2010; Pan & Chiang 2010; Michikoshi & Kokubo 2011; Tiscareno 2013; Hoffmann et al 2015) and observations of Saturn's rings (Sremčević et al 2007; Tiscareno et al 2008; Tiscareno et al 2010; Baillié et al 2013) has shown moons embedded in a disk not large enough to clear a gap form propeller structures. Typically, these small moons or moonlets are at least an order of magnitude smaller than the moons Pan and Daphnis that gravitationally clear gaps which azimuthally extend $2\pi$ radians. The distinctive propeller shape arises from the Keplerian flow of the ring and the gravitational scattering of particles as they pass by the moonlet. In a Keplerian shearing flow the orbital velocity $|\vec{v}|$ of ring particles scales $a^{-1/2}$. Since particles located at smaller radii have shorter orbital periods one arm of the propeller is projected forward and the second arm backwards of the moonlet due to the slower orbital periods of particles with larger semi-major axes.

The origin of J1407b's ring is thought to be the result of a recent collision, leaving a large debris field, analogous to the dynamically young ring system of Saturn which could be due to the tidal disruption of a large object (Dones 1991; Dones et al 2008; Canup 2010; Hyodo et al 2017; Dubinski 2017). However, Saturn's rings and the ring around J1407b are distinctly different since all of Saturn's rings are known to be located within the Roche limit for water ice. J1047b's ring extends well beyond the Roche limit ($\sim 0.001 AU$) and fills a significant portion of its Hill sphere (taken to be $\approx 0.96 AU$ for semi-major axis $a = 5AU$ and the lower mass limit of $20 M_{jup}$ (Rieder & Kenworthy 2016)). Where a large eccentricity is considered for the orbit of J1407b this further reduces to the approximate outer edge of the ring system. A highly eccentric orbit would also result in outer parts of the ring system being located outside the Hill radius when J1407b is at pericentre (Rieder & Kenworthy 2016). A collision between two large rocky objects within the Hill sphere of J1407b could also explain the favoured retrograde orbit of the ring reported by Rieder & Kenworthy (2016). Here it was found that a highly eccentric orbit of J1407b around the primary destabilised a prograde ring while a retrograde ring was able to survive multiple close passes of the primary. It is expected that spectroscopic measurements of the ring during its next transit will identify its orbital direction. These differences between the J1407b ring and Saturn's suggest that active accretion might be present due to its radial location beyond the Roche limit and assumed truncated gaps by unseen exomoons.

Exomoons have recently gained interest in the search for extra solar habitable worlds instead of search for Earth like exoplanets (Quarles et al 2012; Forgan & Kipping 2013), as an extension to the habitable zone occurs through an internal tidal heating (Heller & Barnes 2013; Heller et al 2014). Searches for exomoons are yet to conclusively discover any with current detection methods (Noyola et al 2016; Teachey et al 2017; Forgan 2017; Hwang et al 2018), although one strong candidate does exist (Teachey et al 2018). Therefore, this exoring could offer a viable avenue for extending our knowledge of exomoons and their formation.

## 2. Method

We create a ring of 10,000 massless particles around the exoplanet J1407b. As ring particles are assumed to be massless they evolve only in the presence of J1407b and the moon. For our study we adopt an isolated ring system around J1407b i.e. not orbiting the primary stellar companion J1407. The orbit of J1407b is thought to be highly eccentric with an eccentricity $e > 0.6$ (Rieder & Kenworthy 2016) which we discuss later in the context of this work. It is also assumed that the ring is dusty and contains no significant gas component compared to circumstellar disks (Mannings & Sargent 1997; Andrews & Williams 2007), therefore ring particles are not subjected to any hydrodynamical forces in our models and is typical of planetary rings. Dynamical evolution of ring particles is solely through gravitational interactions with the moon and J1407b.

Integration is performed with the use of Gadget-2 (Springel 2005) and employs a collisionless approach to particle interactions. Gadget-2 uses a smoothing kernel to reduce gravitational forces within a set radius of particles. For J1407b the smoothing length relates to its physical size ($M = 20M_J$, Rieder & Kenworthy 2015) and assuming an internal density like Jupiter, $\rho = 1.31\ gcm^{-3}$. For the exomoon we take an internal density comparable to Earth, $\rho = 5\ gcm^{-3}$ to derive smoothing lengths. However, it should be noted that due to the moon being located outside the ring system in our models, ring particles are unlikely to get close enough to the moon such that smoothing lengths become important.

We place J1407b at the origin of our system of coordinates with a zero-magnitude velocity vector. The equations for initial positions of all ring particles is given as,

$$\boldsymbol{R_{ring}} = [r \cdot \cos\theta,\ r \cdot \sin\theta, 0] \qquad [2]$$

Where $r$ represents the radial position of ring particles and $\theta$ is the angular position (between $0 - 2\pi$) of ring particles around J1407b. We assume a ring with radial parameters $0.2AU < r < 0.6AU$. The moon particle is placed at various locations external to the ring, $a > 0.6AU$ which correspond to the MMR's (2:1, 7:3, 5:2, 3:1 & 4:1) with ring particles located at $r = 0.4AU$. For simplicity we take the eccentricity of the moon particle to be $e = 0$ in all models. The centre of the observed gap with semi-major axis $a = 0.4AU$ equates to an orbital period of $1.83yrs$ when $M_{J1407b} = 20M_{Jup}$. We allowed the system to evolve for 50 orbital periods of ring particles located at $0.4AU$, which equates to ~$90yrs$, but decreases to ~$46yrs$ for the same number of orbital periods when $M_{J1407b} = 80M_{Jup}$. We calculate the semi-major axes of the moon located at MMR's assuming the following equation from the orbital period $\left(P = 2\pi\ \sqrt{a^3/GM_{J1407b}}\right)$,

$$a = \sqrt[3]{\frac{GM_{J1407b}P^2}{4\pi^2}} \qquad [3]$$

Since our models assume a ring with an outer edge at $0.6\ AU$ we simultaneously investigate if moons in MMR's with inner ring particles are also responsible for the inferred truncated edge of the rings (Kenworthy & Mamajek 2015). Assuming a moon mass on the order $M_\oplus$ then the ring edge at $0.6 AU$ is likely to show significant distortion for the 2:1 MMR, which lies at $0.63 AU$, as it is within $3.8 R_{hill}$ (Weiss et al 2009). Note that we only consider MMR's with moons that are not located within the known outer edge of the rings. For example, the 3:2 MMR would place the moon at a semi-major axis $a = 0.485 AU$ which is not consistent with current observations of the ring due to the gap an embedded moon would create.

## 3.     External Moon

After 25 orbital periods of a $0.08 M_\oplus$ moon located at $0.63 AU$ (2:1 MMR) there is significant disruption (Fig 1) to the outer edge at $0.6 AU$. This is comparable to the early stages of the gap clearing process where particles are radially scattered outwards. As expected, particles located at the 2:1 MMR ($0.4 AU$) are perturbed, reducing the number density of particles at this radial location. Scaling up the moon mass we would expect to see the ring edge at $0.6 AU$ to show a greater degree of gravitational scattering. Therefore, we rule out the possibility of a 2:1 MMR with an outer moon being the main cause of the observed gap at $0.4 AU$. The gaps formed at $0.4 AU$ and $0.485 AU$ are still not very prominent even after 50 orbital periods of the ring particles at $0.4 AU$ (~$90\ yrs$ for the lower mass case of J1407b ($M_{J1407b} = 20 M_{Jup}$).

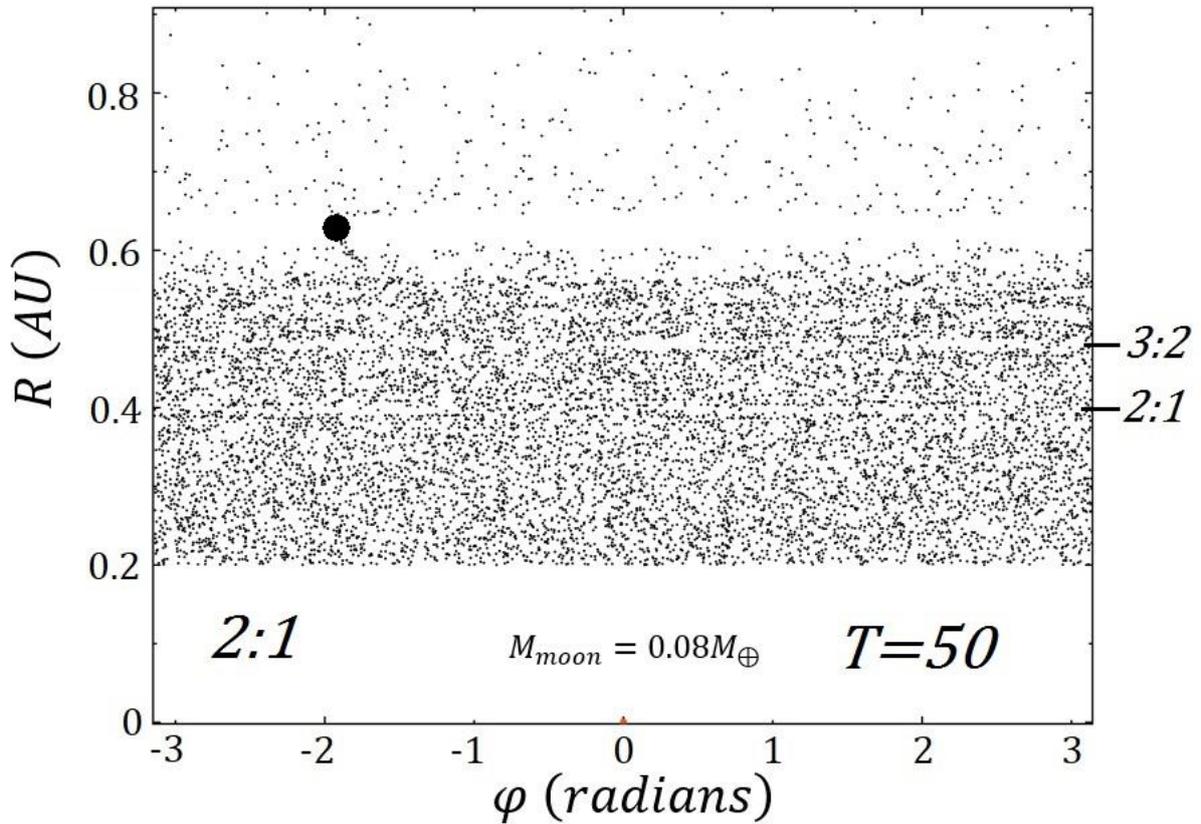

**Figure 1 |** The radial and angular position of ring particles taken at a time of 25 orbital periods of the moon located at $0.63AU$ (2:1 MMR) or 50 orbital periods for ring particles located at $0.4AU$. The mass of the moon is $M_{moon} = 0.08M_\oplus$ and is shown with the black circle. The edge of the ring is located at $0.6AU$ and shows significant gravitational scattering due to the moon being situated only $0.03AU$ away. Due to the proximity of the moon, particles from the edge of the ring are scattered to larger radii $> 0.8AU$. Evidence of the 2:1 and 3:2 MMR's can be seen at $0.4AU$ and $0.485AU$ respectively (labelled on the right hand side of the plot). These are still very small changes in the background density of the ring and unlikely to be significant enough to detect in the lightcurve at $0.4AU$.

When the same mass moon ($M_{moon} = 0.08M_\oplus$) is placed further out at the 3:1 MMR no obvious structures are observed in the ring at $0.4AU$ or distortion of the ring edge for the same time as the 2:1 MMR. Therefore, we increase the mass of the moon to $M_{moon} = 0.3M_\oplus$ for the 3:1 MMR (Fig 2). It should also be noted that the moon masses we use from hereon (Fig 3, 4 & 5) are the largest that can exist without significant disruption of the outer ring edge. Therefore, we use more massive moons when the ring – moon separation increases which correlates to the Hill radii of the moon (Eq 1).

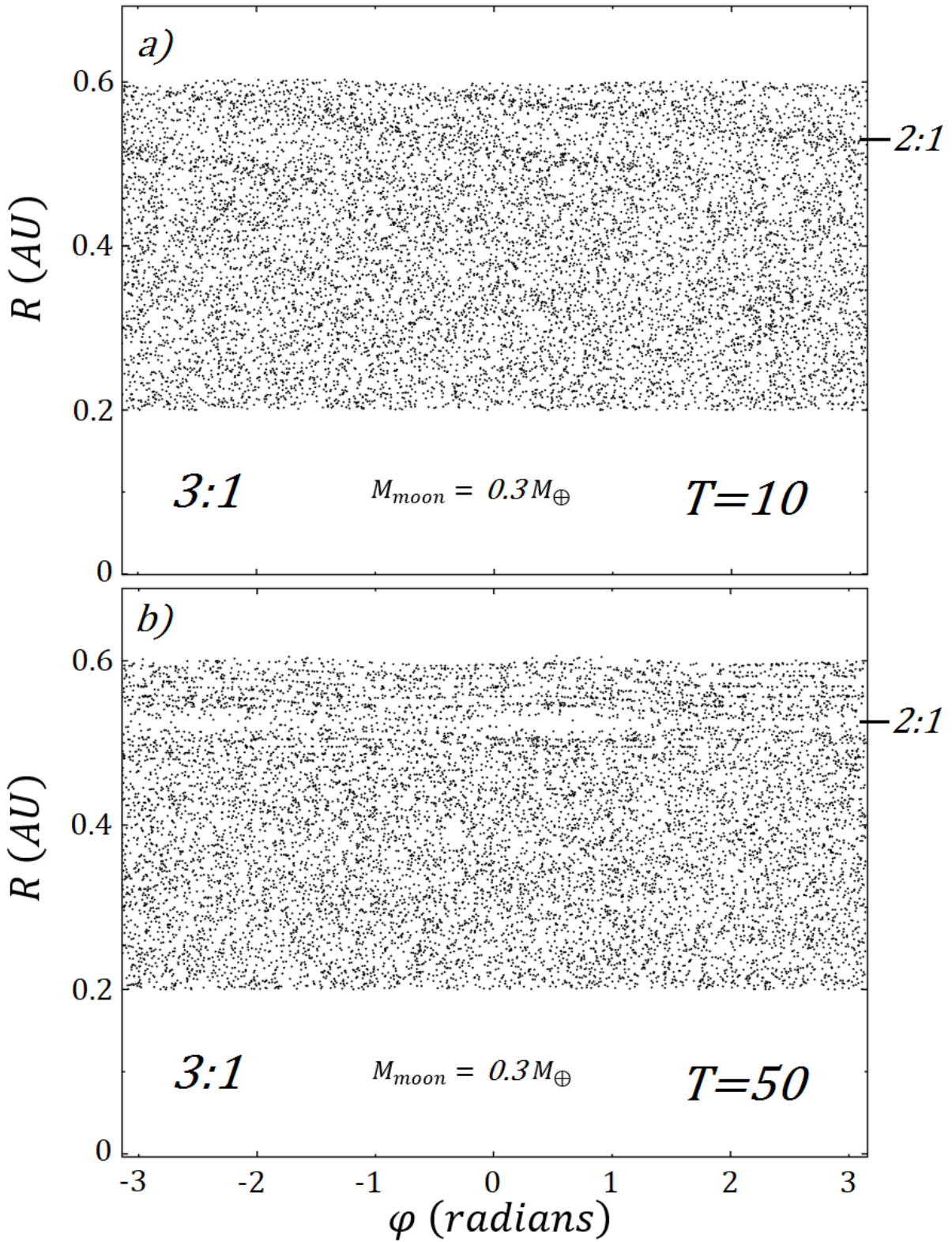

**Figure 2 |** The radial and angular position of ring particles taken at a time of a) 10 orbital periods and b) 50 orbital periods for ring particles located at $0.4 AU$. The moon is located at the 3:1 MMR with ring particles at $0.4 AU$. In this scenario we find that the moon is far enough away that there is no significant distortion or scattering of the ring edge with a moon mass of $M_{moon} = 0.3 M_\oplus$. No obvious perturbation of particles located at $0.4 AU$ is observed during the same time period with only one gap formed at the 2:1 MMR ($0.523\ AU$).

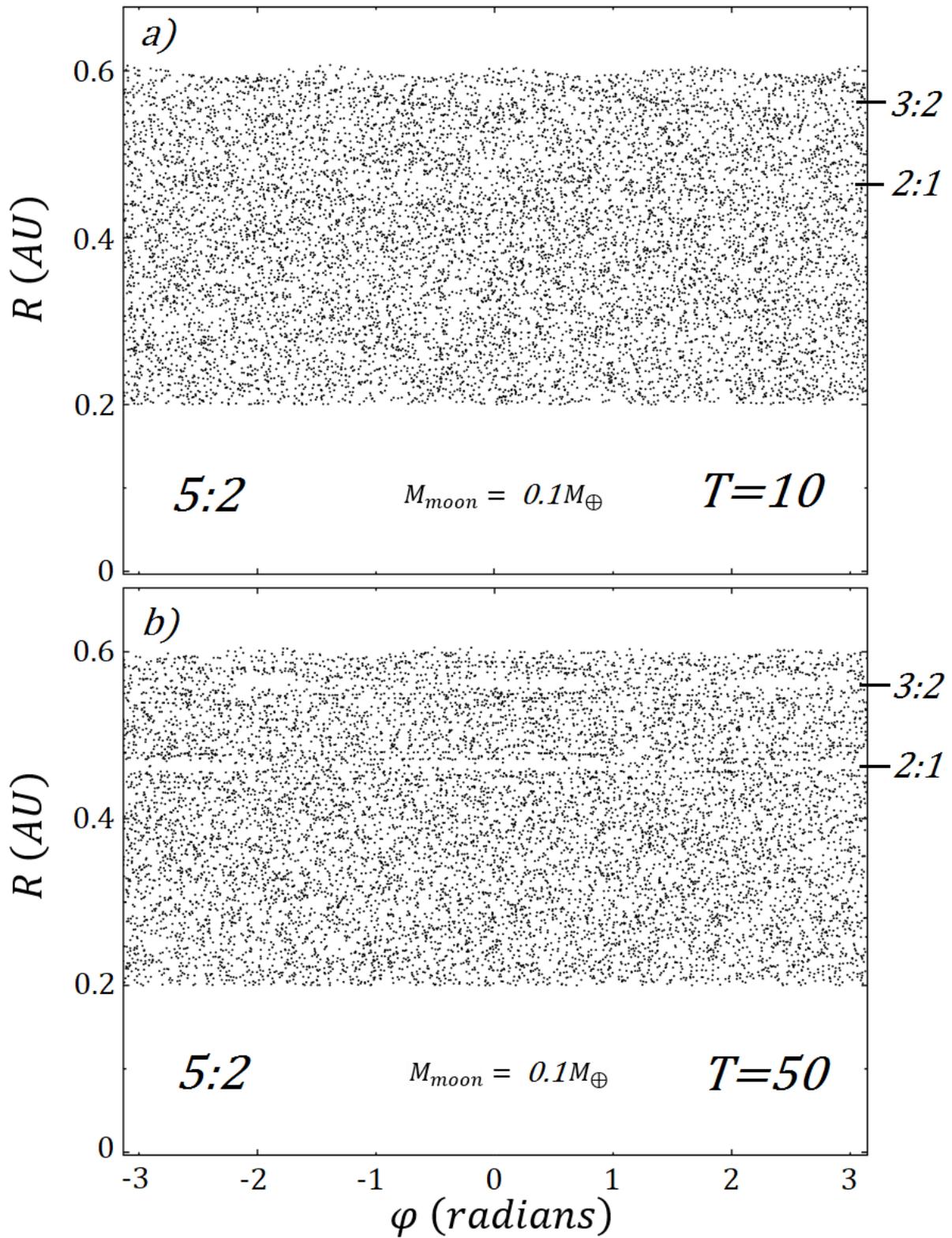

**Figure 3 |** The radial and angular position of ring particles taken at a time of a) 10 orbital periods and b) 50 orbital periods for ring particles located at $0.4AU$ The mass of the moon is $0.1M_\oplus$ and is located at the 5:2 MMR with ring particles at $0.4AU$. For this location of the moon two gap like structures are

observed at ~0.46AU & 0.56AU, which correspond to the 2:1 and 3:2 MMR's respectively (labelled on the right-hand side of the plot). There is no clear evidence of any gap at 0.4AU for the times frames investigated.

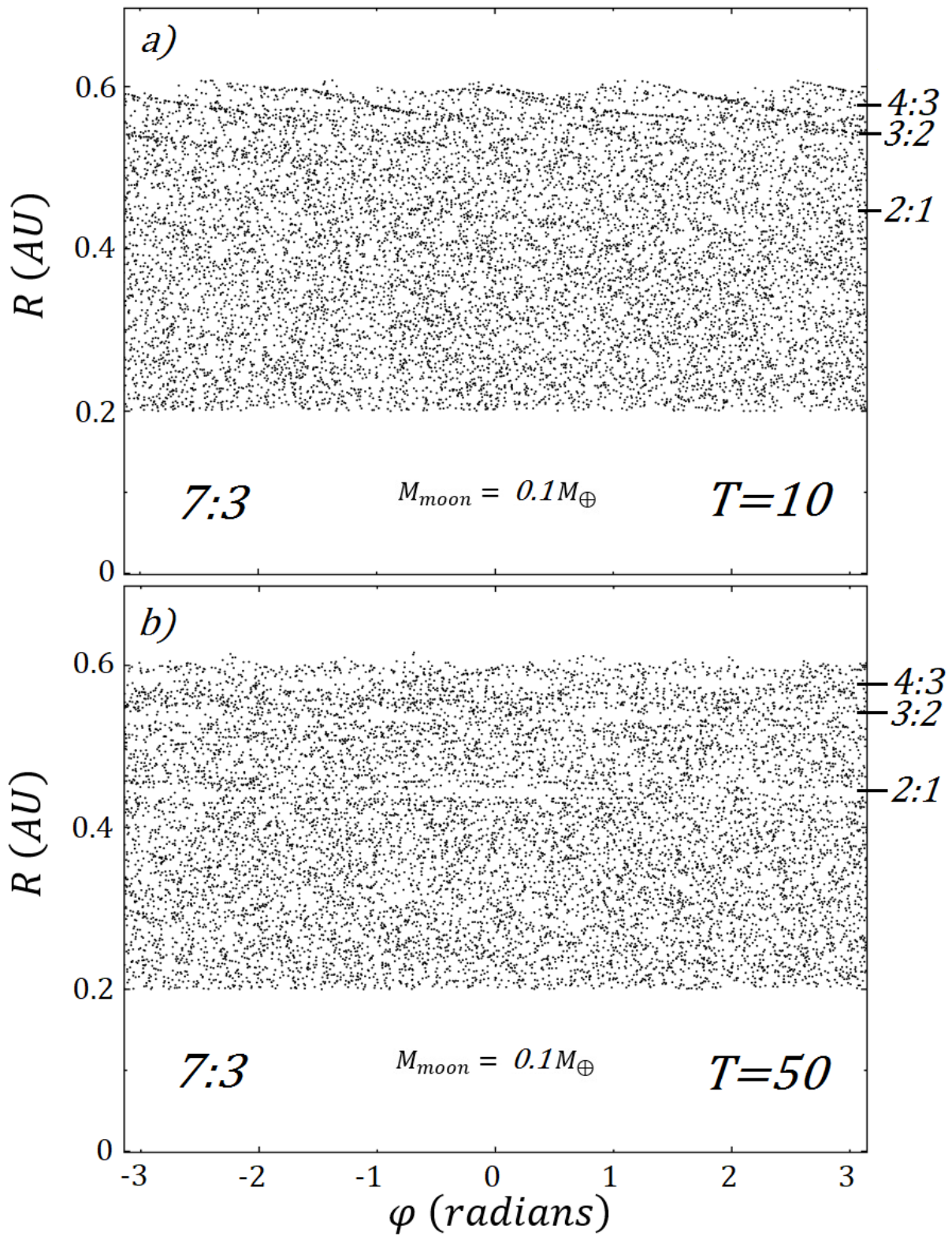

**Figure 4 |** The radial and angular position of ring particles taken at a time of a) 10 orbital periods and b) 50 orbital periods for ring particles located at $0.4 AU$. The mass of the moon is $0.1 M_\oplus$ and is located at the 7:3 MMR with ring particles at $0.4 AU$. For this location of the moon three gap like structures are observed at $\sim 0.443 AU, \; 0.536 AU \; \& \; 0.580 AU$, which correspond to the 2:1, 3:2 & 4:3 MMR's respectively (labelled on the right-hand side of the plot). There is no clear evidence of any gap at $0.4 AU$ for the times frames investigated.

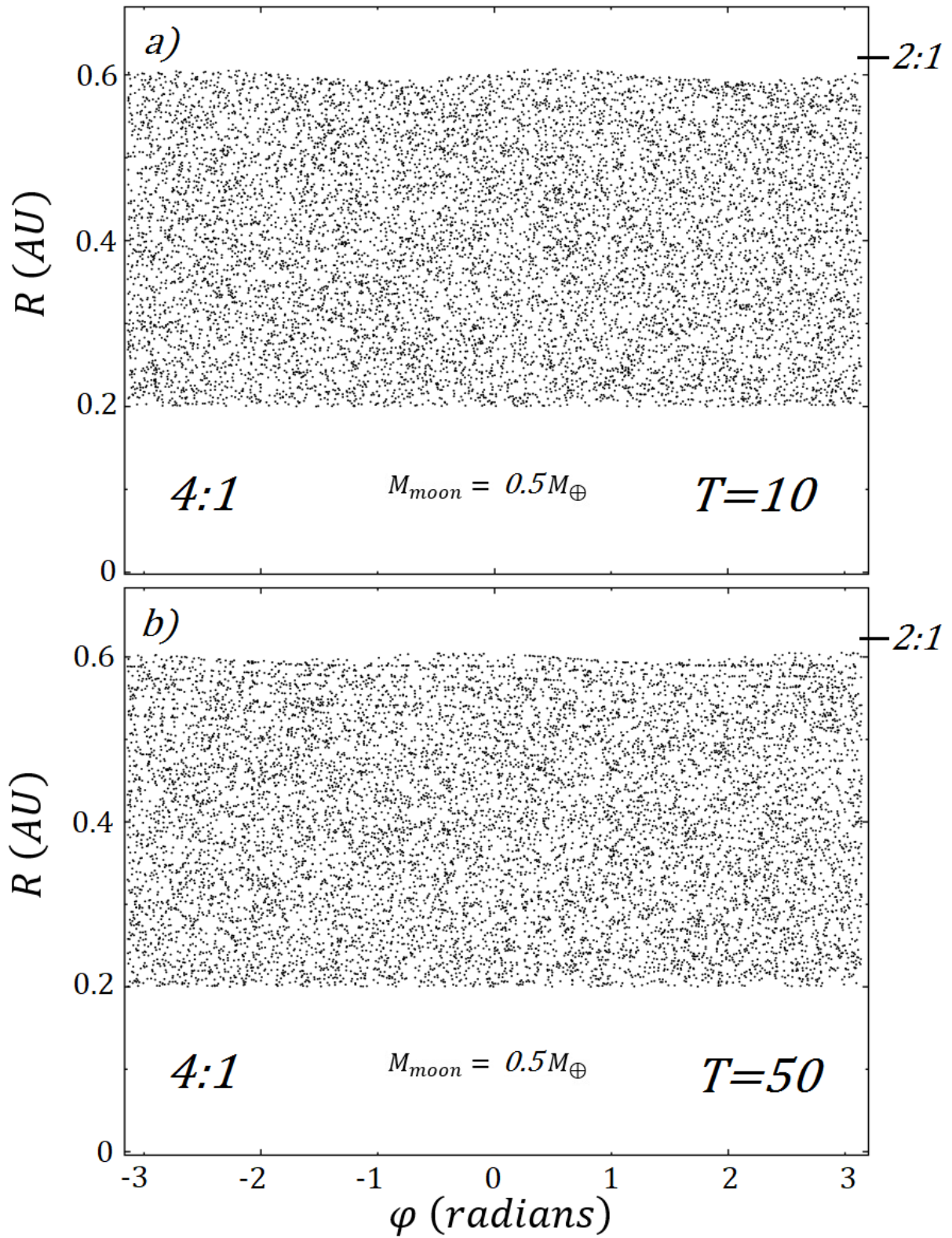

**Figure 5 |** The radial and angular position of ring particles taken at a time of a) 10 orbital periods and b) 50 orbital periods for ring particles located at $0.4 AU$. The mass of the moon is $0.5 M_\oplus$ and is located at the 4:1 MMR with ring particles at $0.4 AU$. For this location of the moon no gap like structures are

observed, however, the ring edge ($0.6AU$) is distorted due to the 2:1 MMR which is located at $0.63AU$. There is no clear evidence of any gap at $0.4AU$ at this time frame.

Modelling of a planetary ring transit by Kenworthy & Mamajek (2015) found that the most significant gap was located at $0.4AU$, yet other small gaps appear on the ring model. Most notable is a $\sim 0.00668AU$ wide gap at $\sim 0.34AU$. Due to the circumplanetary disk nature of the ring (not Saturnian) it is quite possible that these further gaps are caused by embedded moons that have accreted in-situ. Equally, they could be caused by further MMR's with a single external moon at one of the above proposed MMR's. For example, $0.34AU$ lies within $\sim 1\%$ of a 5:1 MMR when an external moon has a 4:1 MMR with the gap at $0.6AU$. However, for the time frames we investigated in our study we do not observe any noteworthy structures at $0.34AU$ & $0.4AU$ when investigating a moon located at the 4:1 MMR. Therefore, it is more probable that embedded moons formed gaps at these locations.

Where a MMR is responsible for creating a gap in the J1407b ring for our time frames ($\sim 90\ yrs$ for the lower mass case of J1407b ($M_{J1407b} = 20 M_{Jup}$) it would need to be either the 2:1 or 3:2 MMR. However, this results in the moon being located too close to the ring edge at $0.6AU$ or inside the ring and would not compliment the lightcurve data. Therefore, we strongly rule out the possibility of MMR's with nearby moons in the J1407b system as the cause of observed gaps.

## 4	Extended Time Periods

When we increase the time period of our models from 50 to 100 orbital periods of ring particles located at $0.4AU$ we still do not observe any gap formations at $0.4AU$. The gaps formed previously (Fig 1 – 5) do not become any more prominent when we extend our models (Fig 6 - 9).

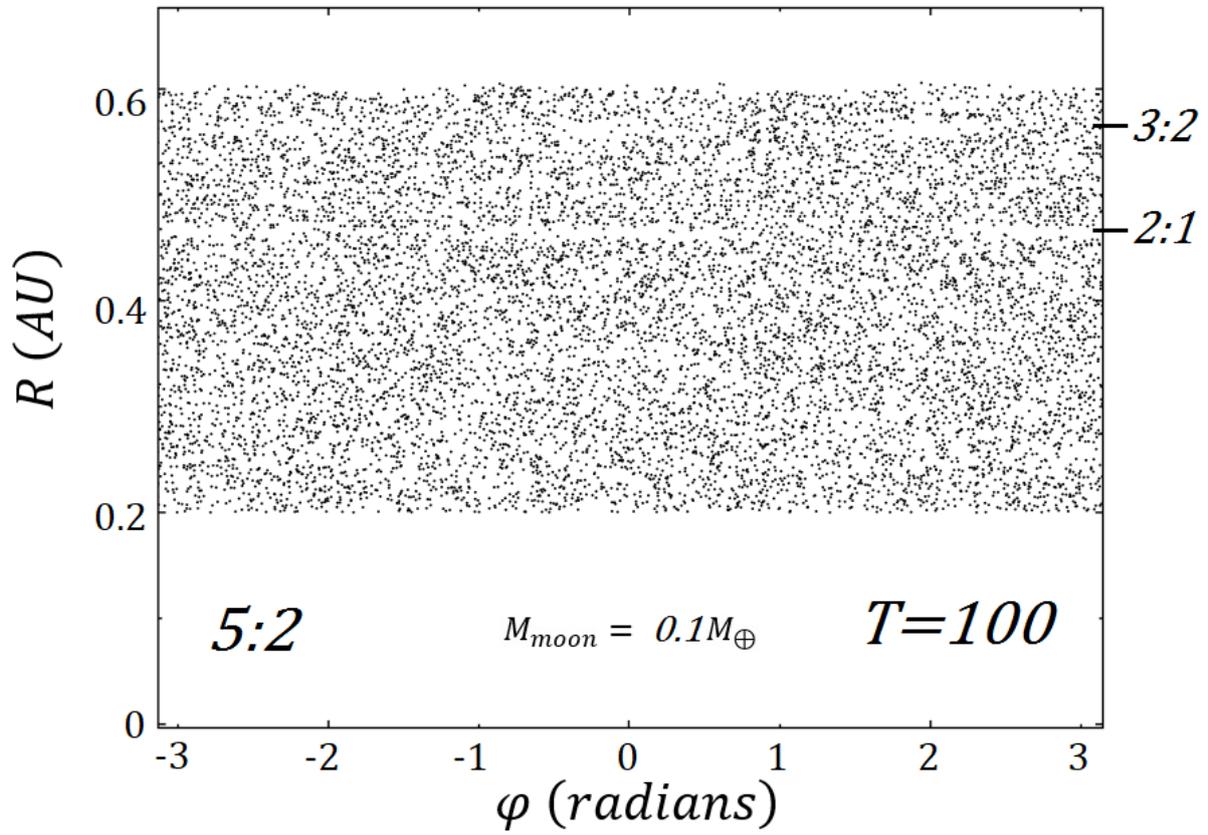

**Figure 6 |** The radial and angular position of ring particles taken at a time of 100 orbital periods for ring particles located at $0.4AU$. The mass of the moon is $0.1M_\oplus$ and is located at the 5:2 MMR with ring particles at $0.4AU$.

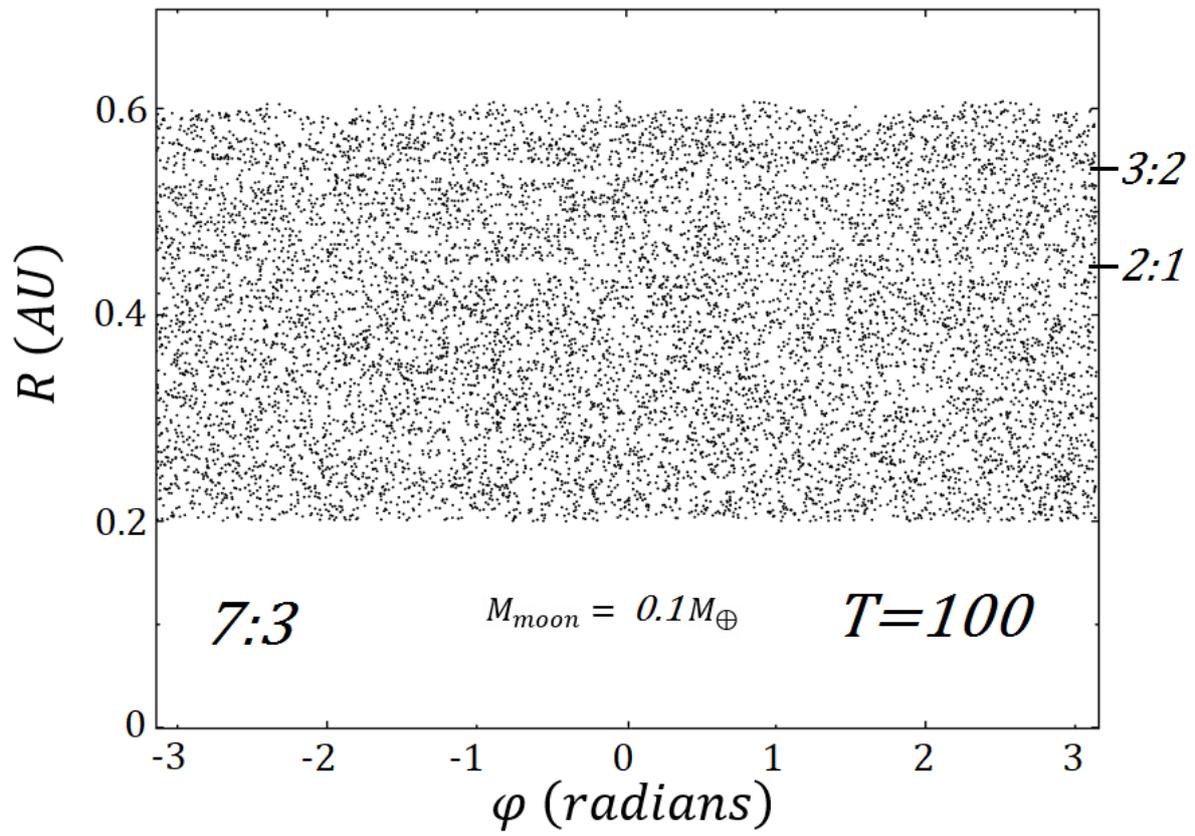

**Figure 7 |** The radial and angular position of ring particles taken at a time of 100 orbital periods for ring particles located at $0.4 AU$. The mass of the moon is $0.1 M_\oplus$ and is located at the 7:3 MMR with ring particles at $0.4 AU$.

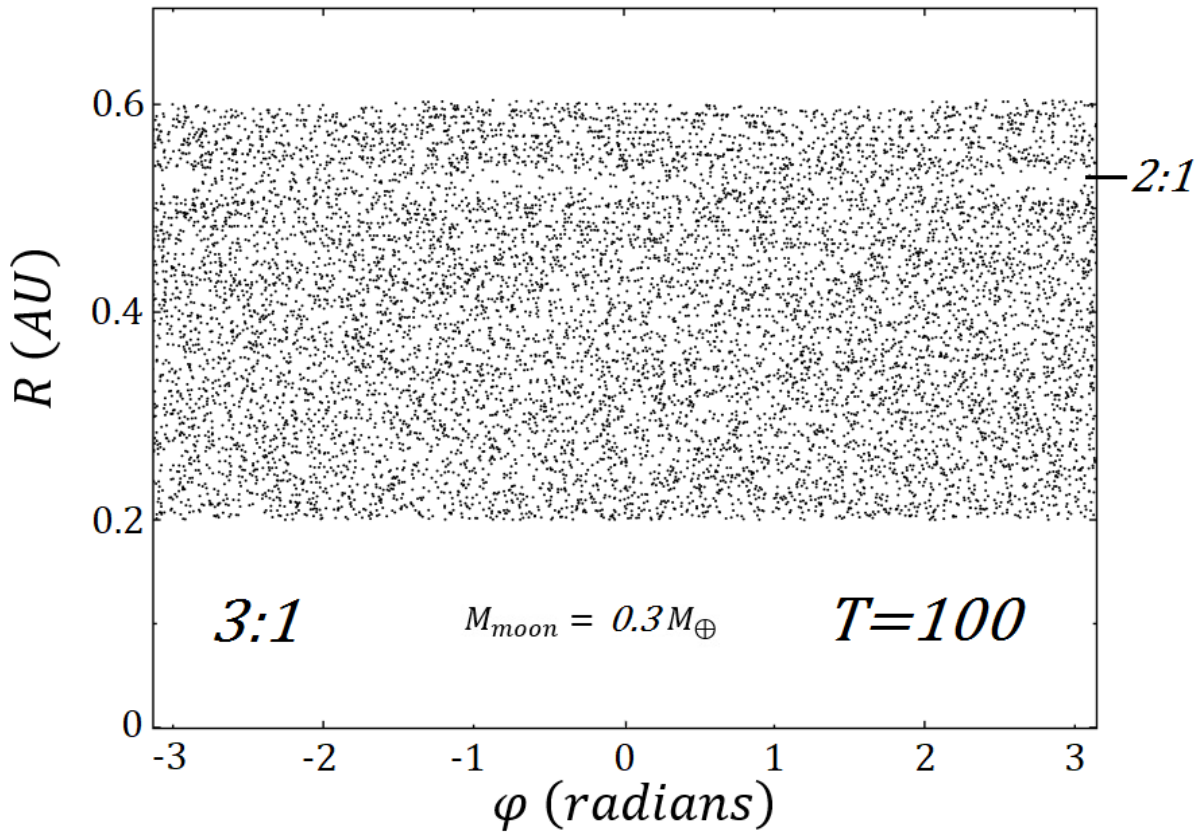

**Figure 8 |** The radial and angular position of ring particles taken at a time of 100 orbital periods for ring particles located at $0.4 AU$. The mass of the moon is $0.3 M_\oplus$ and is located at the 3:1 MMR with ring particles at $0.4 AU$.

## 5. Effects of Ring Mass

The inferred ring around J1407b was estimated to have a total mass of $M_{ring} = M_{Earth}$ (Kenworthy & Mamajek 2015). Therefore, we created a model (moon located at the 5:2 MMR) where the total mass of the ring was $M_{ring} = M_{Earth}$ and comparable to best estimates of the ring mass. In comparison to a ring with no mass we see a slight dampening effect, with gaps less prominent for the same time duration. This is expected since a self-gravitating ring will offer some counteraction to the perturbations of a nearby moon.

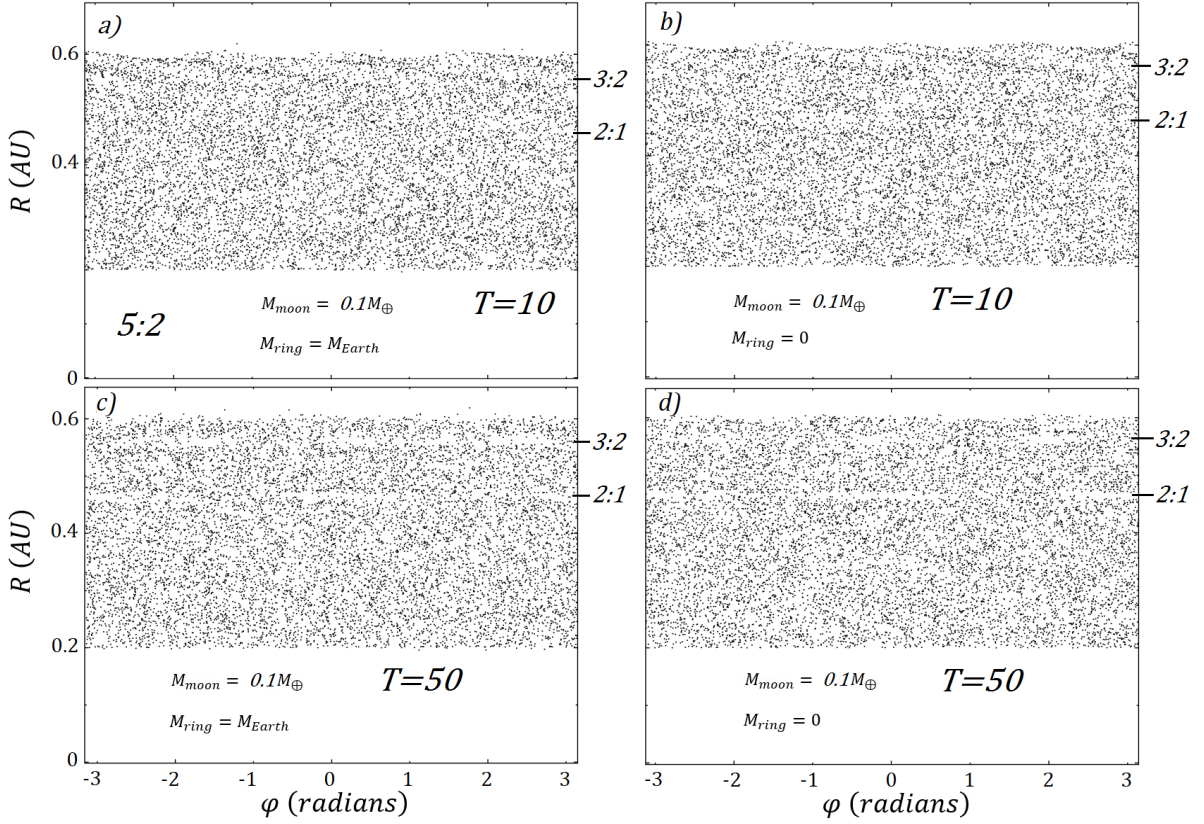

**Figure 9 |** The same model as Fig 3 (b & d) was ran but with a total ring mass of $M_{Earth}$ (a & c). Here, the same gaps are formed but are less prominent during the same time frames. The gaps correspond to the 2:1 and 3:2 MMR's which are located at $\sim 0.46 AU$ & $0.56 AU$ respectively (labelled on the right hand side of the plot). The less prominent visibility of the gaps is due to the dampening effect of self-gravity within the ring as the moon perturbs individual particles.

## 6.  Eccentricity

At radial locations where MMR's occur, we would expect the eccentricity of ring particles to peak since there is an accumulation of gravitational perturbations from the moon. Therefore, we take the magnitude of the eccentricity vector $|\vec{e}|$ and plot it with respect to radial position $R$. Where the eccentricity vector is given as,

$$\vec{e} = \left(\frac{|\vec{v}|^2}{\mu} - \frac{1}{|\vec{r}|}\right)\vec{r} - \frac{\vec{r}\cdot\vec{v}}{\mu}\vec{v} \qquad [4]$$

Figures 10 – 14 show the eccentricity of the same models as Fig 1b – 5b at a time of 50 orbital periods of the particles located at $0.4\ AU$. What we find is that the MMR's become more visible at radial locations previously identified, with the 2:1 and 3:2 MMR's typically being the most dominant. Where the moon is located close to the ring edge (Fig 1 & 10 when located at the 2:1 MMR) the gravitational scattering of particles is clearly seen. Again, the only scenario to show any evidence of a gap or MMR at $0.4\ AU$ is where the moon is placed at the 2:1 MMR with particles at $0.4\ AU$ (Fig 10). All other models where the ring is located further from the ring edge show no evidence of a MMR at $0.4\ AU$ for the time frames investigated in our study. Further MMR's are also observed in the model where the moon is placed at the 2:1 MMR (4:3 & 5:4 in Fig 10) and 7:3 MMR (4:3 in Fig 13) which were not easily identifiable by a gap. These are all close to the ring edge, lying within $0.08\ AU$ of $0.6\ AU$.

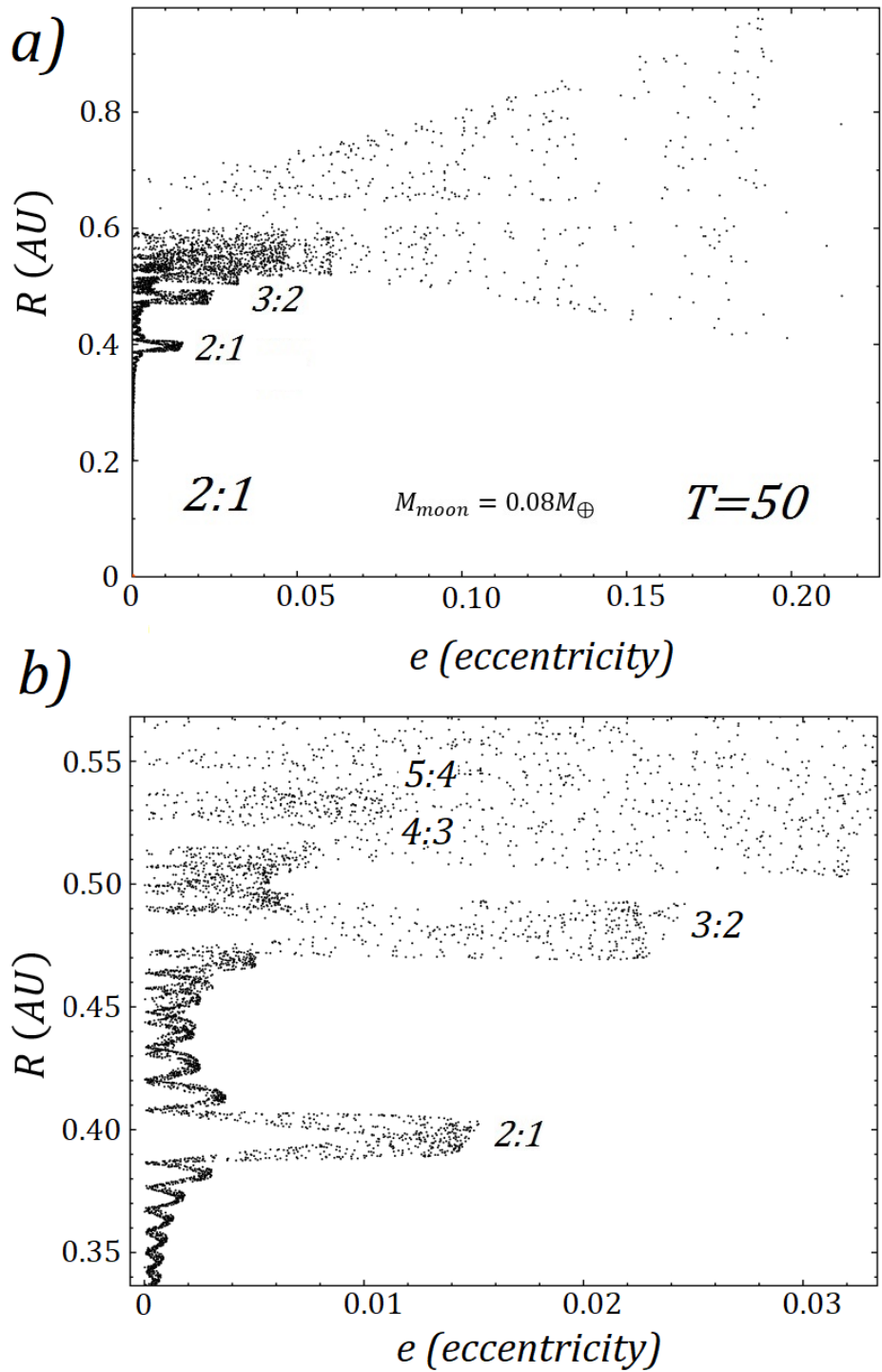

**Figure 10** | The eccentricity and radial position of ring particles is shown at a time of 50 orbital periods of particles located at $0.4\ AU$. This is the same time and model as in Fig 1b. The nearby moon is placed at a 2:1 MMR with ring particles at $0.4\ AU$. a) Here, the two main MMR's are clearly seen at 2:1 and 3:2, along with the scattering of the ring edge due to the proximity of the moon. b) Zoomed in section of (a)

shows there is evidence for at least a further two MMR's located close to the ring edge and the moon which are noted as 4:3 and 5:4 at 0.523 & 0.546 $AU$ respectively.

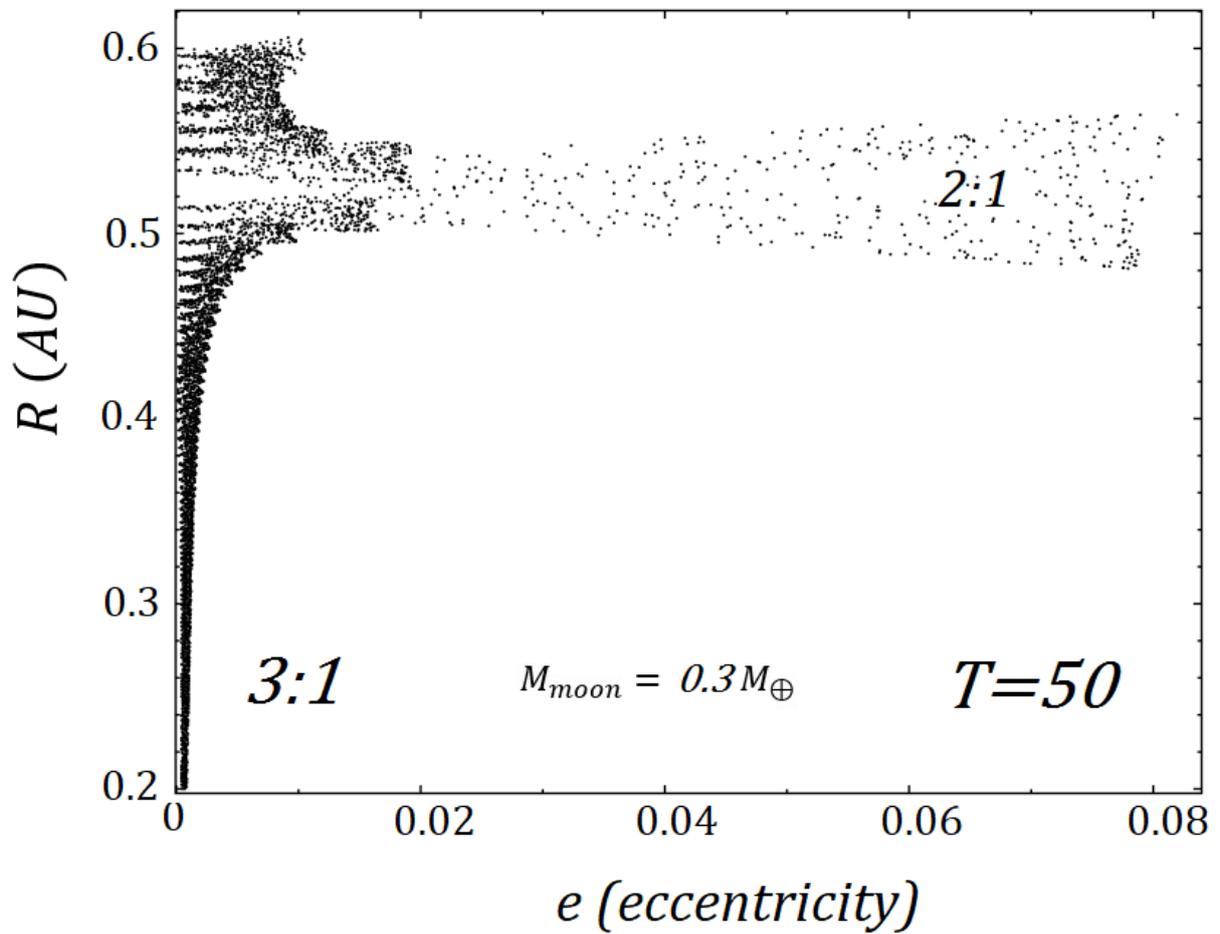

**Figure 11 |** The eccentricity and radial position of ring particles is shown at a time of 50 orbital periods of particles located at $0.4\ AU$. This is the same time and model as in Fig 2b. The nearby moon is placed at a 3:1 MMR with ring particles at $0.4\ AU$. There is only evidence of the 2:1 MMR located at $0.523\ AU$ in this plot.

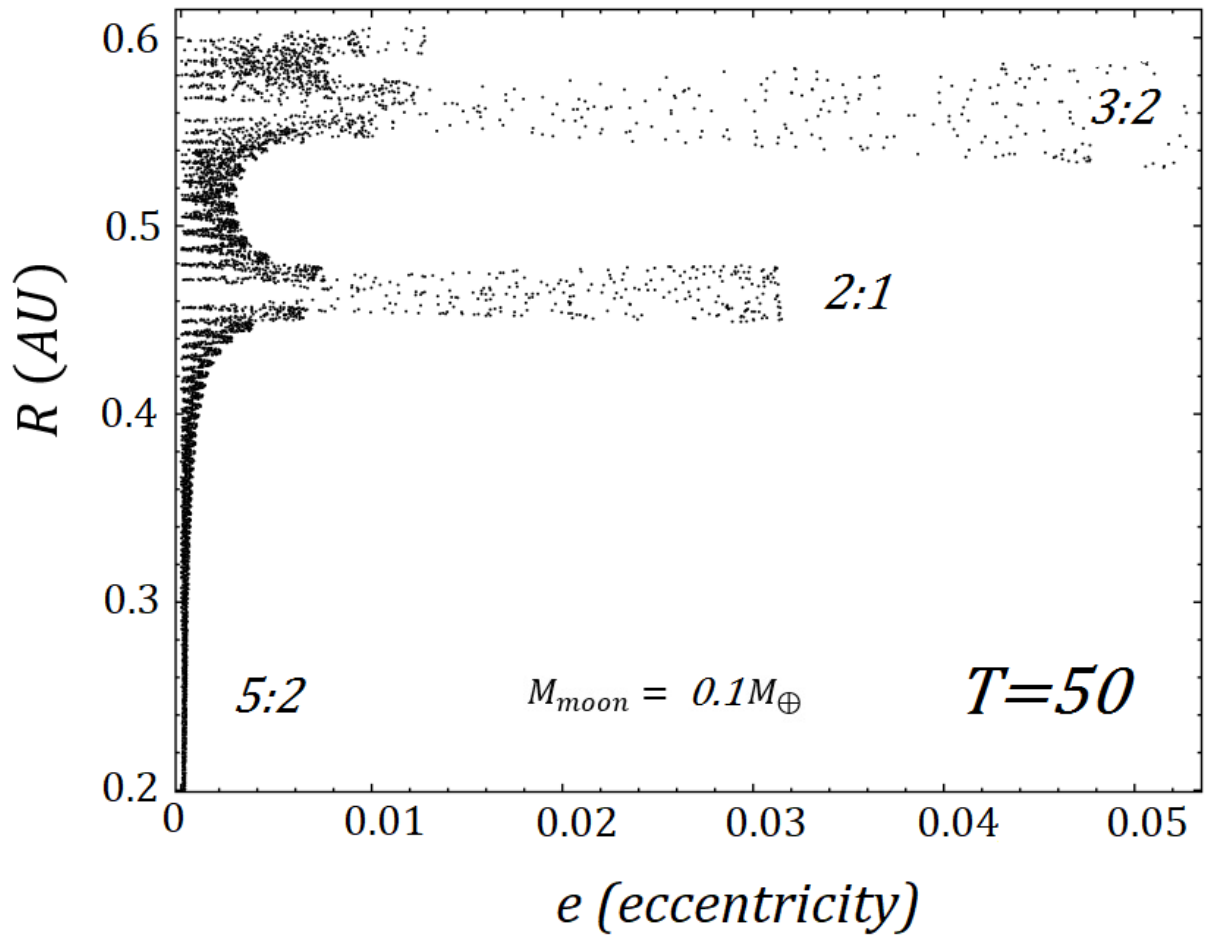

**Figure 12 |** The eccentricity and radial position of ring particles is shown at a time of 50 orbital periods of particles located at $0.4\ AU$. This is the same time and model as in Fig 3b. The nearby moon is placed at a 5:2 MMR with ring particles at $0.4\ AU$. Here, there is evidence of the 2:1 and 3:2 MMR located at $0.46AU$ & $0.56AU$ in this plot.

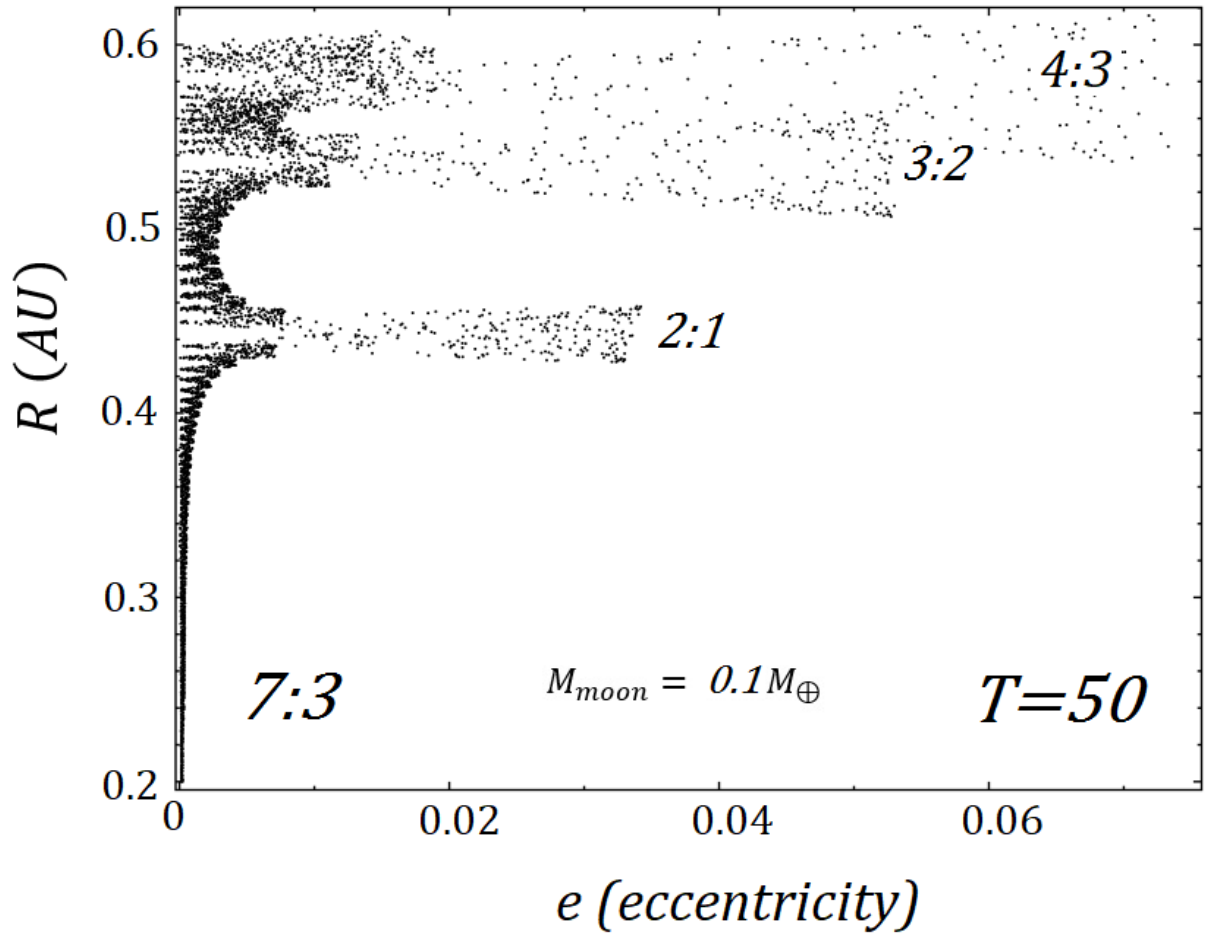

**Figure 13 |** The eccentricity and radial position of ring particles is shown at a time of 50 orbital periods of particles located at $0.4\ AU$. This is the same time and model as in Fig 4b. The nearby moon is placed at a 7:3 MMR with ring particles at $0.4\ AU$. Here, there is evidence of the 2:1, 3:2 & 4:3 MMR's, which are located at $0.443 AU,\ 0.536 AU$ & $0.580 AU$.

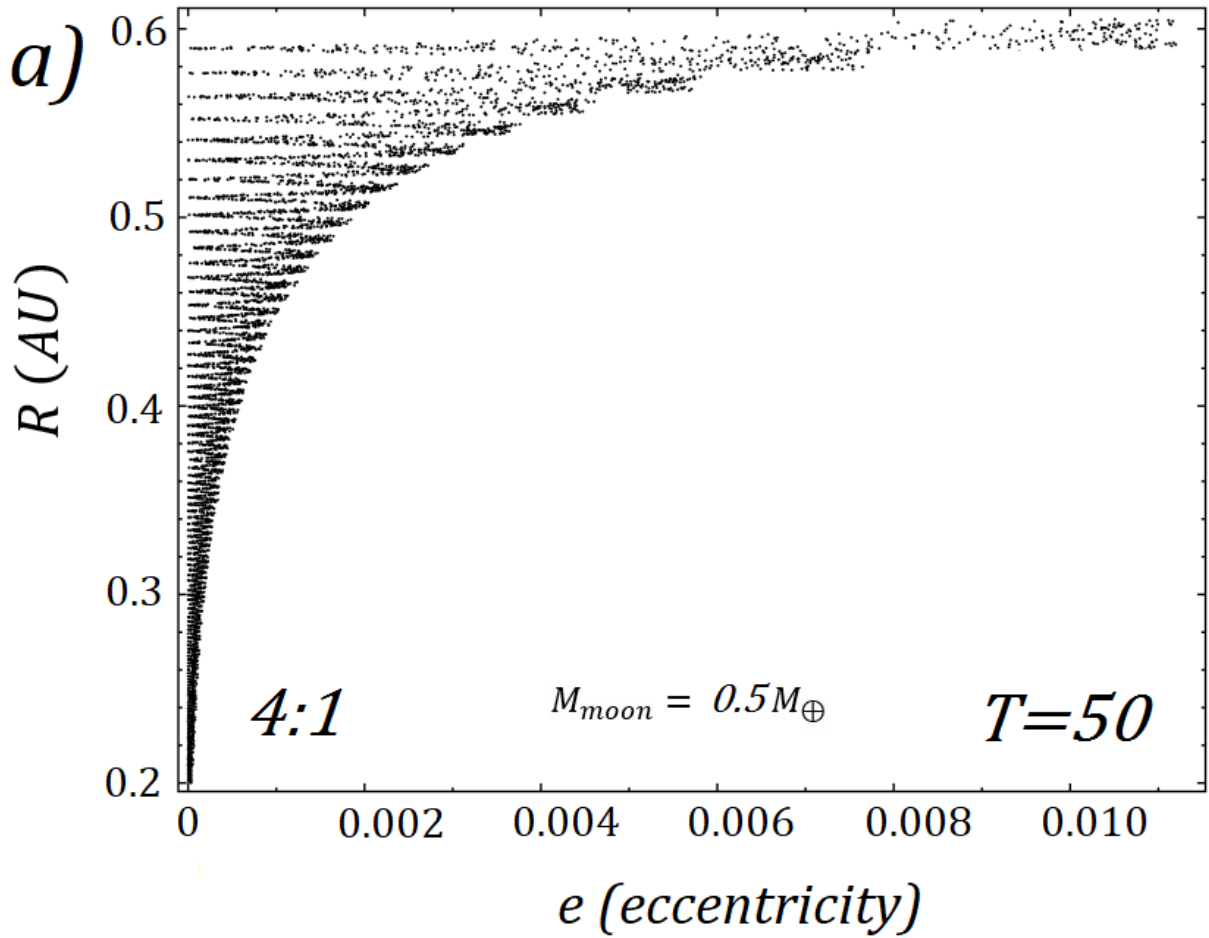

**Figure 14 |** The eccentricity and radial position of ring particles is shown at a time of 50 orbital periods of particles located at $0.4\ AU$. This is the same time and model as in Fig 5b. The nearby moon is placed at a 7:3 MMR with ring particles at $0.4\ AU$. Here, there no evidence of any MMR's, except some distortion of the ring edge which occurs to due to the 2:1 MMR being located just outside the ring.

## 7. Discussion

The main purpose of this study was to investigate the gap opening at a radial location of $0.4AU$ in the assumed ring around the exoplanet J1407b due to MMR's with nearby moons. By placing an external moon with masses $0.08M_\oplus < M_{Moon} < 0.5M_\oplus$ at the MMR's 2:1, 7:3, 5:2, 3:1 & 4:1, we conclude that it is unlikely that MMR's with nearby moons are the cause of the observed gap at $0.4\ AU$ in the J1407b ring system. Nevertheless, embedded moons still offer a potential mechanism to form the inferred gaps that were observed in the lightcurve as opposed to external moons in MMR's. Only the moon with a semi-major axis of $0.63\ AU$ (2:1 MMR) forms a gap at $0.4\ AU$. However, another notable gap is also formed in this scenario at the 3:2 MMR (Fig 1) at $0.485\ AU$ which would have been detectable on the original lightcurve. Secondly, the proximity of the moon causes a significant scattering of the ring edge at $0.6\ AU$, beyond the semi-major axis of the moon, which again is not observed in the transit of J1407b.

Photometric analysis of J1407 indicates a total ring mass off $M_{ring} \cong 1.23 M_\oplus$ (Kenworthy & Mamajek 2015). We showed that for one of the models (5:2) incorporating a total ring mass of $M_{ring} = M_\oplus$ caused a dampening effect on the gap formation process due to the self-gravity of the ring. Therefore, future work should incorporate a self-gravitating ring with an appropriate total mass and some consideration to collisions between ring particles, which would likely alter and extend the time frames of gap formation due to MMR's with external moons (Lewis & Stewart 2009). Although global models incorporating such additional physics would come at a significant computational cost. The relatively low number of particles in our models means collisions between particles would be very low, where considerably larger global simulations are performed particle collisions will then become more important.

We should also note that we cannot directly compare our results with the Cassini Division in Saturn's rings, which lies at a 2:1 MMR with the moon Mimas. The moon does indeed play a role in causing some of the structure in the Cassini Division and apparent gap. However, much of the finer scale structure is still not understood and has no apparent association with perturbing satellites (French et al 2010; Hedman et al 2010; French et al 2016). Thus, care should be taken when comparing the Cassini Division and our models.

Caveats to our work:

1. We do not consider collisions in our models, which has been shown to alter the gap formation process in planetary rings. This causes a diffusive effect downstream of the moon-ring perturbations and results in gaps filling back up with particles. Thus, in a system that considers collisions we would expect gap formation to be impeded and time scales for observable gaps to increase.

2. We do not consider the highly eccentric orbit of J1407b about the primary which is thought to take the ring system close enough at pericentre that the Hill radius reduces below the measured radial extent of the ring system (Rieder & Kenworthy 2016).This would cause significant distortion of the ring, making gap formation by MMR's very difficult given the time frames we investigate and the estimated orbital period of J1407b of $\sim 11\ yrs$. Furthermore, non-negligible eccentricities would be introduced to the moon and ring particles. For embedded moons the width of the gap they form in a ring scales with the Hill radii $\left(R_{Hill} = a(1-e)\sqrt[3]{M_{moon}/3M_{J1407b}}\right)$. The Hill radii of embedded moons also decreases with eccentricity $e$ due to the decrease in separation between the moon and planet at pericentre. The embedded moons in Saturn's rings, Pan and Daphnis, were shown to create gaps with half widths of $\Delta a \approx 3.8 R_{Hill}$ (Weiss et al 2009), but only where $\Delta a$ was small compared to $a$ and where ring particles encounter the moon on circular orbits. However, the gap opening process of a highly eccentric embedded moon has not been properly investigated and could deviate away from the less eccentric case of Saturn's embedded moons where $\Delta a$ remains small compared to $a$. Due to uncertainties on dampening and stability post perturbation the idea of an embedded moon forming a gap within a disrupted ring is still an open question.

3. If J1407b is bound to the primary on a highly elliptical orbit then moons external to the ring system ($> 0.6 AU$), which are required for MMR's, will exist outside the Hill radius during pericentre. Thus, raising serious concerns on the stability of any moons outside the current ring system.

Consequently, future work should look at the effect of external moons and ring interactions during the close encounters of J1407b with the primary, and if the gaps are still able to form between each orbit of J1407b. Intuitively this does not appear plausible since it takes multiple orbits to form a fully cleared gap ($> 20\ yrs$), which is greater than the time it takes for J1407b to orbit the primary ($P_{J1407b} \sim 11\ yr$). However, there is still considerable uncertainty on the semi-major axis, eccentricity and orbital period of J1407b due to the lack of a second transit. In addition to the gap formation process during close encounters at pericentre, future work should also investigate gravitational instabilities in the ring and subsequent moon formation. Are close encounters between J1407b and the primary beneficial or detrimental to the accretion of moons within the ring? This could be analogous to Saturn's F ring where a nearby moon Prometheus disrupts the F ring causing local gravitational instabilities and subsequent small moonlets, which were found by the Cassini spacecraft (Murray et al 2005; Beurle et al 2010; Sutton 2018). In contrast, more recent work (Mentel et al 2018) suggests that J1407b might not be bound to J1407, and if true would be beneficial for both scenarios of gap formation by MMR's of nearby moons and embedded moons since the ring would not be strongly perturbed.